\documentclass[letterpaper,USenglish]{lipics-v2019}

\nolinenumbers
\hideLIPIcs  

\usepackage{url} 
\hypersetup{colorlinks=true,linkcolor=red,urlcolor=blue,citecolor=blue}

\usepackage{tabularx} 
\usepackage{booktabs}

\usepackage[maxfloats=150]{morefloats}                                               
\usepackage{microtype}                                                               
\usepackage[utf8]{inputenc}   
\usepackage{courier} 
\newcommand\Invisible[1]{                                                            
  \marginpar{\color{white}{\fontsize{.5}{.5}\selectfont #1 }}                        
}  

\newcommand{\Exclude}[1]{}

\newcommand\Boldly[1]{\vspace{0.5 \baselineskip} \noindent \textbf{$\blacktriangleright$} \textbf{#1} \noindent}
\newcommand\BoldSection[1]{\vspace{0.5 \baselineskip} \noindent \textbf{#1} \noindent}
\definecolor{Gray95}{gray}{0.95}
\definecolor{forestgreen}{rgb}{0.13, 0.55, 0.13}

\newcommand\RedZone[1]{
  \begingroup\color{red}{#1}\endgroup
}

\newcommand{\AtFoot}[1]{\let\thefootnote\relax\footnotetext{{#1}}} 

\newcommand{\remove}[1] {}

\usepackage{microtype}   %if unwanted, comment out or use option "draft"
\usepackage{listings} 
\usepackage{graphicx} 
\usepackage{amssymb} 
\usepackage[scaled]{beramono} 
\usepackage{bera} 
\usepackage[T1]{fontenc}
\usepackage{changepage} 

\usepackage[iso,american,cleanlook]{isodate}                                         
\usepackage[export]{adjustbox} 
\usepackage{fancyhdr}                                                             
\makeatletter\let\ps@plain\ps@fancy\makeatother
\usepackage{background}
\backgroundsetup{
  angle=0,  
  scale=1, 
  opacity=1,
  firstpage=true,
  position=current page.south west, 
  hshift=84pt,
  vshift=460pt,
  contents={\colorbox[rgb]{0.99,0.78,0.07}{Page \thepage}}
}

\makeatletter
\lst@Key{countblanklines}{true}[t]%
    {\lstKV@SetIf{#1}\lst@ifcountblanklines}

\lst@AddToHook{OnEmptyLine}{%
    \lst@ifnumberblanklines\else%
       \lst@ifcountblanklines\else%
         \advance\c@lstnumber-\@ne\relax%
       \fi%
    \fi}
\makeatother

\lstdefinestyle{numbers}
{numbers=left, stepnumber=1, numberstyle=\tiny, numbersep=10pt}
\lstdefinestyle{nonumbers}
{numbers=none}

\usepackage{tikz} 
\usepackage[inline]{enumitem}

\usepackage{bm} 

\usepackage[sc]{mathpazo} 

\usepackage[ampersand]{easylist}

\usepackage[margin=15pt,font=small,labelfont={bf,sf}]{caption}

\title{Compact Java Monitors} 

\author{Dave Dice}{Oracle Labs}{dave.dice@oracle.com}{https://orcid.org/0000-0001-9164-7747}{}
\author{Alex Kogan}{Oracle Labs}{alex.kogan@oracle.com}{https://orcid.org/0000-0002-4419-4340}{}
\authorrunning{D. Dice and A. Kogan} 

\Copyright{Oracle and/or its affiliates}


\category{}%optional, e.g. invited paper

\relatedversion{A full version is available at \url{https://arxiv.org/0000.00000}} 

\supplement{}%optional, e.g. related research data, source code, ... hosted on a repository like zenodo, figshare, GitHub, ...

\funding{}%optional, to capture a funding statement, which applies to all authors. Please enter author specific funding statements as fifth argument of the \author macro.

\acknowledgements{We thank David Holmes, Doug Lea and Erik Österlund for discussions on early drafts.
Peter Buhr, Thierry Delisle and Chris Parsons (Univserity of Waterloo) also reviewed early drafts
and provided helpful comments.} 

\EventShortTitle{}

\begin{document}

\maketitle

\begin{abstract}
For scope and context, the idea we'll describe below, Compact Java Monitors, is intended as a potential
replacement implementation for the ``synchronized'' construct in the HotSpot JVM.  
The readers is assumed to be familiar with current HotSpot implementation. 
\end{abstract}

\Invisible{Oracle Invention Disclosure Accession : ORC2133570-US-PSP} 

%% Would strongly prefer to use proper header/footers ..
%% \AtFoot{\today \hspace{1mm} $\blacklozenge$ \hspace{1mm} Copyright Oracle and or its affiliates} 
%% \AtFoot{\today \hspace{1mm} \textbullet \hspace{1mm} Copyright Oracle and or its affiliates} 

\section{Introduction}

\textbf{Compact Java Monitors (CJM)} are based on the Compact NUMA-Aware Locks (CNA) algorithm, 
but ignoring the \emph{NUMA-Aware} property and focusing on the \emph{Compact} aspect.  CNA is itself a 
variation on the gold-standard MCS (Mellor-Crummey Scott) \cite{tocs91-MellorCrummey} queue-based lock algorithm 
\footnote{If you're unfamiliar with MCS, please see the appendix for a quick description.}.
%% From page one of \url{https://arxiv.org/pdf/2003.05025.pdf}   

%% \symbol{"200B} 

Underlying much of the following design is our approach from Compact NUMA-Aware Locks 
(CNA) which was published in EuroSys 2019 \cite{EuroSys19-CNA,arxiv-CNA}  and is 
being integrated into the Linux kernel as a replacement for the existing low-level
\emph{qspinlock} construct \cite{linux-locks,Long13}, which is itself based on MCS.  
CNA is itself a variation on classic MCS.  One of the key ideas in 
CNA is propagating values of interest from the MCS owner's queue node into the successor,
which allows the lock body to remain compact -- just one word.  
Specifically, fields that would normally appear in the body of a lock are
instead maintained in the owner's queue node and, at unlock-time, conveyed to the successor in the queue. 
In the context of this discussion we're not interested in NUMA-aware aspects of CNA, where we 
propagate the list of remote nodes isolated on a distinct chain,  but instead
the fact that the lock is compact.  Taken to the extreme, our design shifts \emph{all}
the fields that'd normally reside in the classic HotSpot \emph{objectMonitor} construct 
into the MCS queue nodes, so we can represent the abstract monitor with just a single pointer 
to the MCS tail.

\section{Requirements for \texttt{synchronized}} 

Lets start with requirements for \texttt{synchronized} in Java.  Taking some liberties, the
API is basically \texttt{lock} (\texttt{MONITORENTER}), \texttt{unlock} (\texttt{MONITOREXIT}), 
\texttt{wait}, \texttt{notify} and \texttt{notifyAll}.  
As \texttt{notifyAll} is usually just a trivial variant of \texttt{notify}, we'll 
use the term \texttt{notify} to collectively refer to both flavors.  
For reasons I'll explain shortly, the identity \texttt{hashCode} facility is also intimately convolved with
synchronization, as they share the \emph{mark} word (described below), so any subsystem redesign needs to 
take that into account.  There are also JNI (Java Native Interface) equivalents for the above, which also 
need to be supported.  The language demands that locking operations expressed via \texttt{synchronized} 
need to be lexically balanced
\footnote{There is no way to express imbalanced locking in the language itself, and 
the \texttt{javac} compiler will only emit bytecode that is balanced.  
Bytecode verification does \emph{not}, however, require balanced locking, but the
C2 JIT checks for balance, in order to assign stack slots for locking constructs,
and any code that fails the balance check is banished to run in the interpreter.}. 
JNI locking operators, which mirror those 
above but are callable from C/C++, are slightly more liberal, but not to the point
of any particular complication.  Note that lexically balanced locking precludes such
common idioms as hand-over-hand (coupled) locking.  Relatedly, the only way to acquire
an arbitrary number of locks is via recursion, which imposes its own constraints because
of stack growth.  Empirically, from papers which characterize synchronization 
behavior, we believe that it is extremely rare for a thread to hold more than 3 
distinct locks at any given time.  

The \texttt{unlock, notify} and \texttt{wait} operators all require that the caller hold the lock
in question, otherwise they're required to throw ``Illegal Monitor State Exception'',
abbreviated here as \texttt{IMSX}.  Conceptually, you can think of Java monitors as 
melding a \texttt{pthread} condvar and mutex, but you can only use the condvar while holding
that associated mutex.  Recursive locking is required, so if you acquire a lock and then
acquire it again, that's fine, and the JVM will track the recursion levels.  
(Recursive locks make sense in an object-oriented framework where some synchronized 
methods might call other publicly exposed synchronized accessor methods).  The JIT 
is often able to detect recursive locking and discard the inner synchronization.  
JIT-based automatic inlining helps in this effort.   It's also worth noting that 
since lock and unlock operations must be lexically paired, we can often relax the 
ownership check in the unlock operation.  As JNI operations have not such 
language-level lexical constrains, we always must check ownership for JNI 
operations.  

Our desiderata are the usual.  We want good performance, which means low latency absent
contention and good scaling under contention.   
We assume a spin-then-park waiting model, made loom-aware by omitting the spin phase as
appropriate.  Indefinite spinning or yielding isn't acceptable.  
We also need to provide \emph{wait morphing}
to accelerate \texttt{wait/notify}
\footnote{\emph{Wait morphing} is a performance optimization where \texttt{notify} simply moves
threads from the waitset to the \emph{entryset} -- the MCS chain in our case -- avoiding
the need to immediately wake the notifyee at the point where \texttt{notify} executed. 
This avoids futile context switching where the notifyee might otherwise contend on the lock still
held by the thread that called \texttt{notify}, and, critically, may shift expensive 
unpark operations -- often thousands of cycles -- out of the critical section.  
HotSpot's current implementation provides wait morphing.  The linux \texttt{pthread\_mutex} 
and \texttt{pthread\_condvar} implementation also provides wait morphing for 
\texttt{pthread\_cond\_signal} and \texttt{pthread\_cond\_broadcast}, when the runtime
can establish that the signalling thread holds the same mutex specified by the waiter(s).}. 
Fairness is more interesting.  The current implementation allows unbounded bypass.  
(It's worth noting that all the \texttt{Java.util.Concurrent} (JUC) constructs do as well, except
one flavor of ReentrantLock where \emph{FIFO} is explicitly selected).  
For loom, given that MCS provides direct handoff, we
can directly context switch the carrier thread to the successor, which usually
provides good performance and responsiveness.  We also care about less tangible 
qualities, such as simplicity.  The current system is extremely complex, 
having evolved through accretion, which inhibits research and experimentation.  
Ideally, a new subsystem would be more malleable and comprehensible.   In addition, 
the current subsystem is ``marbled''\cite{BigBallOfMud}  and tightly or overly coupled to other subsystems -- 
disentangling such dependencies is highly desirable.  We also want the lock and 
related structures to be compact to control footprint.  

%% \section{HotSpot and some background on the implementation of \texttt{synchronized}} 
\section{HotSpot : background} 

Briefly, HotSpot had, until recently, 3 locking modes : \emph{biased locking}; 
\emph{stack locking}; and \emph{inflated locks}.  Biased locking attempted to address the issue 
of expensive atomics.  If a given lock was acquired and release repeatedly by just 
one thread, then, conceptually, we'd defer the unlock until contention arose at 
which point we'd release the lock.  This optimization -- which is a latency play, 
not a scalability improvement -- made economic sense when atomics were expensive 
(particularly in the time of bus locking, before cache-based atomics) but changes 
in modern architectures have sped up CAS, etc., making biased locking obsolete and of little value.  
It was also complicated and made the code far more brittle.  Thankfully, 
biased locking has been recently removed, leaving us with stack locking and 
inflated locks.  Stack locking is typically used in uncontended operation.  
At the onset of contention, we'll \emph{inflate} an object and switch to inflated mode.  
(There are other reasons we inflate as well, but those aren't really relevant to this
discussion).   

\Invisible{sclerotic} 

At this point we need a quick digression into the object header.  In HotSpot each 
object has 2 header words.  These words are accessible by the runtime, but not by 
application code.  The 1st word is the so-called \emph{class pointer}, which describes 
the type of the object.  This is used for certain runtime casting checks, for 
garbage collection (GC) to determine the type and layout of the object for root processing, and, if 
necessary, it also contains information equivalent to a C++ vtable pointer.  
The 2nd word is the \emph{mark} word, which is the focus of our interest.  
The mark is heavily overloaded --
a single-word discriminated union, where the low-order bits act as a tag and describe what 
the other bits contain.  The mark can variously hold the 
identity hashCode value, a pointer into the owner's stack for stack locking mode, 
or a pointer to an inflated object monitor when in inflated state.  In the latter 
two cases the hashCode value is said to be \emph{displaced} and stored elsewhere.  HotSpot 
assigns hashCodes on-demand, lazily, but once a hashCode is assigned to an object, 
the relationship must be stable and permanent for the lifetime of the object.  
It's worth noting that the object header format is not dictated in any specifications 
and is at the whim of the implementer.  The heavily overloaded and multiplexed 
mark word is a HotSpot-specific design decision.  Since \emph{every} object in the heap 
has a header, there's pressure to keep the header size as small as possible.

Back to synchronization, the stack locking mode isn't particularly interesting 
except to say that it adds complication and clutter, requiring triage of the 
mark word low-order bits and requiring the ability to shift from stack locked to 
inflated state upon contention.  When inflated, the high order bits of the mark 
point directly to a native C++ \texttt{objectMonitor} structure.  The objectMonitor, or 
simply ``monitor'' or ``fat lock'', contains a lock; the identity of the owning 
thread, if any; recursion level; waitset pointer; a back-pointer to the associated 
object; and a list of threads waiting on entry.  An objectMonitor can be referred 
to and be associated with at most object at any given time.  

Until recently, in HotSpot, once an object was
inflated, it would stay inflated until at least the next safepoint.  That is, the 
relationship between an object was stable except at certain points.  This simplified
the design, as once a thread has fetched from the mark and found a pointer to an 
objectMonitor, it can depend on that relationship holding unless the code passes 
through a safepoint.  In some incarnations of HotSpot, the objectMonitors resided 
in type-stable memory (TSM).  The downside to this design is that we can suffer 
rampant objectMonitor accumulation.  It's not uncommon to have 10s of thousands of 
extant objectMonitor instances, reflecting a significant footprint.  Until recently 
deflation only occurred only at safepoints -- monitor scavenging -- which is 
undesirable, as safepoints are driven (for the purposes of this discussion) by 
GC activities.  Specifically, deflation at safepoints is expensive, not highly 
parallel, and reflects an unwanted coupling between the GC and safepoint mechanism, 
and the synchronization subsystem.  Recent versions of HotSpot have eliminated 
safepoint-time monitor scavenging and deflation, and deflate outside safepoints,
partially addressing the issue of monitor accretion and footprint. 

\subsection{Historical Perspective : ExactVM} 

\Invisible{Received knowledge; lore; folkore; wisdom; historical; folk myth} 

\Invisible{Evolution of HotSpot metalock implementation} 

It's worth mentioning that some JVMs deflate aggressively on the ``last'' unlock, 
when there are no waiting threads.  Sun's ExactVM used this strategy.  While appealing, 
this adds extra synchronization to avoid races between arriving threads and the 
deflator.  Specifically, lets say object $O$'s mark refers to monitor instance $M$.  
At unlock-time, the owner does not observe any contending or waiting threads and 
wants to deflate and retire $M$ -- severing the linkage between the mark word and $M$ --
while some other thread is concurrently arriving to lock the object and has just 
read that mark word in the object header and is about to follow that pointer to $M$.
Techniques are available to either prevent or detect and recover from this race, but they all 
involve extra synchronization and interlocks in the fast uncontended path, impacting
performance.  ExactVM used a simple test-and-set bit -- 
the \emph{metalock}\cite{oopsla99-agesen} -- in the mark word that threads 
had to acquire when enqueuing or deflating, so ultimately, under very high 
concurrent lock flux, we just shifted all the queueing and contention to that bit 
\footnote{ExactVM's metalock is an example of an \emph{inner lock}, which protects
internal locking metadata such as queues.  At high arrival rates, 
contention on the logical outer lock results in the inner lock becoming contended 
and acting as a bottleneck.  
For instance, contention in \texttt{phread\_mutex\_lock} on linux often translates
into contention and waiting on the kernel's \texttt{futex} per-bucket chain locks.}. 
Also, the arrival and deflation paths were longer because of the more involved protocol on that bit.  

Furthermore, with aggressive early ``ASAP'' deflation, common producer-consumer idioms 
cause monitors to flow or migrate from one set of threads to another, presenting 
yet another memory management challenge.  To improve scalability, ExactVM allocated 
monitors in bulk, moving monitors from a global free list to per-thread free lists.  
Each time a thread needed more monitors we'd double the allocation unit size for 
that thread, clamped to 64K monitors, and at safepoints we'd reset the bulk 
reprovisioning size for each thread back to 16, and also trim those local free lists 
that had accumulated excessive numbers of free monitors.

To reduce coherence traffic on the central lock-free free list, which proved
to be a bottleneck, we added a \texttt{home} field to monitors to indicate their
origin and added a lock-free per-thread free list where the associated thread could 
push and pop, and remote threads could only push, allowing monitors that had 
migrated between threads to be returned to their origin.  
This presumes that the direction of inter-thread monitor ``flow'' will persist and is not random,
which, empirically, seems to be the case.  That is, previous outflow predicts future outflow.  
The extra overhead imposed by lock-free operations added latency, however,
so we instead provided each thread with a purely local free list as well as a lock-free
\emph{returning} list.  In the unlock operator, if a thread deflated and recovered 
a foreign monitor, it would use CAS to push that monitor onto the lock-free remote 
\texttt{returning} stack associated with the monitor's home thread.
To allocate a monitor for inflation in the lock operator, a thread 
would first check its private local free list.  If the local list was empty it 
would then revert to the \texttt{returning} list, using \texttt{SWAP} of \texttt{null} 
to detach the set of monitors en masse, and then move that string of 
recovered monitors back to the local free list.  
Failing all the above, the thread would resort to allocation from the global pool.  
ExactVM did not have thin locks or stack locks, so an
initial uncontended lock operation needed to allocate a monitor and inflate,
so reducing the latency of allocation was critical.  

\Invisible{So I then switched to a local list that was only ever accessed by 
the associated thread, and a new ``returning'' list that was a stack of nodes 
pushed with CAS, and detached by SWAP by the home thread.  
If its local list became empty, a thread would try to detach the whole returning 
list and move that to its local list, otherwise it’d have to fall back to 
allocate from our process-wide global list.} 

\Invisible{We ultimately added a \texttt{home} field to monitors indicating their origin.  If a thread
deflated and recovered a foreign monitor, it would push that monitor onto a 
lock-free remote return stack associated with the monitor's home thread.
This presumes that the direction of monitor ``flow'' will persist and is not random,
which, empirically, seems to be the case.  That is, previous outflow predicts future outflow.  
Threads had a primary private list of free monitors, but, if the list became empty, 
would also check their remote return list -- detaching it en masse via a SWAP of 
\texttt{null} -- before resorting to allocating more monitors from the global pool.  While 
this provided some relief, it still was not a complete solution and we'd still end
up with more extant monitors in circulation than wanted. } 

See Appendix-\ref{Appendix:lifecycle} for a comparison on legacy HotSpot, modern HotSpot, ExactVM and CJM. 

\section{Compact Java Monitors : construction -- simple locking} 

We'll now describe, incrementally, a different approach.  Initially, we'll ignore 
hashCodes, and \texttt{wait/notify}, and start out with classic MCS assuming we have the whole 
object mark word at our disposal to serve as a pointer to the MCS tail.  
Note that we now forego a classic objectMonitor, replacing that construct with a 
chain of MCS queue nodes.  MCS is frugal and parsimonious, requiring only one queue 
element (``node'') for each lock a thread holds and one additional node if the thread 
is waiting, yielding a very tight bound.   A simple per-thread free list of nodes 
suffices, grown on-demand and implemented as stack.  (More on this below).  
Nodes are always returned to the free list of the thread that allocated and 
enqueued the node, avoiding rampant footprint growth or inter-thread migration, 
as described above.  There should be no need to trim free lists
\footnote{If the need to trim arose, the implementation is trivial and a thread-local decision.
Absent trimming, a thread's free list plus active list will contain $N+1$ elements where $N$
is the maximum number of locks held at any given time by that thread, putting a tight
bound on the number of extant elements.}. 
MCS also deflates aggressively when the last thread departs, but without any complex 
or expensive mechanisms, such as HotSpot's current deflation sentinel approach or 
ExactVM's metalock.  Recursive locking is tracked with a new \texttt{nesting} field in 
the owner's queue node. 

We also need the ability to efficiently determine if the current thread $T$ has 
locked object $O$ to implement IMSX checks in \texttt{unlock, wait} and \texttt{notify}.  
and to check ownership for recursive locking, and finally we must have 
a way for the unlock operator to quickly find the implicit head (owner) 
of the MCS chain, in order to provide succession and to reclaim that node.  
To that end, we augment the synchronization subsystem with a per-thread associative 
map from objects to installed MCS queue nodes.  (It's worth noting that some 
MCS implementations use a non-standard lock-unlock API and simply pass the 
installed queued node address from the lock site to the unlock site.  Other 
implementations preserve the usual lock-unlock API but store the owner's queue 
node address in the lock body itself, to convey the address from unlock to unlock.  
None of those options are readily available to us in HotSpot).  

In practice, our associative map can be implemented as a trivial linked list of 
active queue nodes, arranged as a stack rooted in thread-local storage.  
\emph{Active} in this case means enqueued on an MCS chain.  Given the stack-like 
nature of \texttt{synchronized}, absent JNI, at unlock-time, the relevant queue node 
will be found at the top of the stack.  We expect accesses via the map to 
be mostly constant-time.  The queue nodes need to be extended to include an 
object reference field to validate the association.  As such, those 
references must be strong GC roots and scanned for any garbage collection 
at safepoints.  While undesirable, we expect the lists to be extremely short
and note that threads are already visited for root processing.  
We can easily integrate the per-thread lists of free queue nodes and held nodes.

\section{Compact Java Monitors : construction -- \texttt{wait} and \texttt{notify}}
 
Arguably, we now have Java-style lock and unlock implemented, but the design sketched 
above does not support \texttt{hashCode} or \texttt{wait/notify}, so we'll now move on
to address the latter. A thread can return from \texttt{wait} because (a) it was notified, 
(b) the timeout, if any, expired, or (c) pending interrupt exceptions.  
In CJM, threads blocked in \texttt{wait} continue to be represented by the queue node they used to 
originally obtain the lock.  The waitset of an object (monitor) is a pointer to a 
linked list of such waiting queue nodes, and is stored in the owner's queue node 
in a distinct \texttt{waitset} field.  Borrowing heavily from CNA, the value of this field -- 
and thus the waitset -- is propagated from the owner to the successor at at unlock-time.  
When a thread calls \texttt{wait}, it must hold the lock, in which case it 
adds itself to its own waitset, installs that waitset into the successor on the 
primary MCS chain, and passes ownership to the successor.  When a thread calls 
\texttt{notify}, it simply transfers threads from the waitset to the MCS chain, 
appending the element(s) with a single SWAP, providing efficient wait morphing.  
The lock itself protects the waitset from concurrent accesses -- only the current
owner can access the waitset.  More precisely, other than appending new threads to 
the MCS chain upon arrival, only the current owner can edit the MCS chain or the waitset.  
It's worth noting that the waitset field is only set in the head (owner's) 
queue element.  Non-head elements will always have their waitset field set to \texttt{null}.  

The case where a thread calls \texttt{wait} and the MCS chain is otherwise empty -- no 
successors exist on the chain -- is subtle and deserves special attention.  
In this circumstance we leave that waiting thread at the head of the MCS chain
as a placeholder but mark its status (a field in the queue node) accordingly.  
Threads subsequently enqueuing on the chain via lock must inspect the predecessor's 
state, and, if it was waiting, take immediate ownership of the lock and shift that predecessor, 
and its waitset, if any, to the new owner's queue node waitset.  We call this 
usurping the lock.   
The special placeholder state is the only circumstance where a thread blocked in \texttt{wait} appears 
resident on the primary MCS chain, and it must be the only element on the chain, 
except briefly, when new threads arrive.   

\Invisible{There is a minor race between threads switching to the 
special waiting state and arriving threads that might usurp ownership, but it's a 1-vs-1 
race and easily managed by using CAS to coordinate and interlock on the status field.  
(We might also encode the special \emph{waiters only} placeholder state via a dedicated low-order
bit in the mark word, avoiding the use of the queue node state field). 
}

Threads also exit wait state via interrupts or timeout, presenting a more interesting 
challenge.  The procedure to implement \texttt{notify} was simple, relatively, as the waitset 
was protected by the lock itself and the notifying thread simply moved the notifyee 
from one list to another.  
For interrupts and timeout, a thread will return from park and notice either the 
timeout or the interrupt condition and must take direct actions itself to address  
the situation.  Assuming the thread is on the waitset (and not in special waiting 
state), the thread then allocates and enqueues a 2nd MCS element -- which we call 
the \emph{beta} element --  onto the MCS chain to acquire ownership.  
The thread then waits until either the original element \emph{or} the beta makes its way 
to the head of the MCS chain and becomes the owner, at which point the thread 
removes the \emph{other} element from either the waitset or the primary MCS chain, and 
finally returns.  This step, of expunging the other node, may involve  
traversal operations on the MCS chain, and requires waiting for the \texttt{next}
MCS links to resolve to non-\texttt{null}.  

\section{Compact Java Monitors : construction -- \texttt{hashCode}}

To recap, at this point we have lock, unlock, \texttt{wait} (with interrupt and timeout 
support) and \texttt{notify}.  Next on the agenda is to add identity \texttt{hashCode} support.  
We'll employ the usual approach of a displaced mark word (DMW).  We require only one 
\emph{tag} bit for the discriminated union in the mark. The mark word encoding 
is $0:0$ for neutral initial state, $H:1$ for hashed and unlocked, where $H$ is the hashCode 
value and 1 indicates the least-significant bit is set, and $N:0$ indicates that the 
object is locked and $N$ points to the tail of the MCS chain in the expected fashion. 
To make things simple, we assign a hashCode to an object either on the first call to \texttt{hashCode} or 
the first time we lock an object.  Generating a hashCode value is cheap.  
%% Similar to CNA, and the way in which we propagated the waitset, above,  
We we store the DMW -- containing the hashCode value -- in a dedicated field 
in the MCS queue elements, and propagate the hashCode value through \emph{all} 
elements of the MCS chain.  
That is, all elements of the MCS chain will carry the same DMW (hashCode) value.

Fetching or assigning the hashCode when the mark is in $0:0$ or $H:1$ state is trivial, 
and similar to the scheme HotSpot uses today.  We use CAS to assign and change from 
$0:0$ to $H:1$ state, if necessary.  If the object is locked, however, then the 
protocol become more interesting.  As called out above, locked implies hashed, 
which simplifies the design.  If the object is locked by the caller, we can simply 
reach through the per-thread map to find the owner's queue node and access the 
DMW to fetch the hashCode.  (Note that I've ignored the mark word's GC age bits, 
but we'll assume for the moment those can be handled, effectively, as part
of the hashCode).  

When the object is locked by some other thread then we need to use a more involved 
strategy to safely retrieve the hashCode value.  In a sense, we're trying to effect
a consistent read of $object\rightarrow{}mark\rightarrow{}DMW$
even though the mark word is subject to concurrent flux from threads arriving 
and departing on the MCS chain.  I'll sketch out one crude but perfectly workable 
variation.  We provide a global array of locks, indexed by a hash on a queue node's 
virtual address.   Readers of the hashCode (assuming the object is locked by 
another thread) will read the mark word, acquire the associated lock, validate 
that the mark still refers to the same object, fetch the DMW, drop the lock, and 
returned the fetched DMW, which must encode a hashCode value.  If validation fails, 
we just drop the lock and loop, retrying.   Complementary code acquires and 
immediately releases the lock when we release a queue node back to the per-thread 
free lists.  We could also defer that action until the node actually recycles.  
We might also use reader-writer locks instead of simple locks but there's likely 
no benefit given that we expect this case to be fairly rare.  

\Invisible{Chase-and-capture or ratify idiom}  
\Invisible{Optimistic hashCode readers; readers-do-no-write; progress properties; obstruction free} 

For the purposes of explication we described a method that uses locks, but we could 
also use an external array of reference counts, hashed by the element address, to pin
a element to allow safe reading of the hashCode.  Using an external array 
of locks or reference counts allows us to avoid type-stable memory of the
elements.  The reference counts act as a simple protection counter incremented
and decremented with atomic fetch-and-add.  This approach is likely the better for a real implementation.
Reading the hashCode is then obstruction-free and threads attempting to recycle a
node would need to wait for the value to be observed as 0.  

As noted above, we require that the hashCode value be propagated through 
the entirety of the MCS chain, via the DMW field in the queue nodes, to allow 
potential readers to fetch the hashCode from tail queue element.  Assuming contended 
locking and a non-trivial chain, a thread arrives and swaps its queue node $N$ 
into the MCS tail, obtaining the address of the predecessor queue node, $P$.  Before 
installing $N$, the locking thread sets $N$'s DMW value to 0 to indicate not-yet-present.   
After the atomic swap, but before setting $P$'s forward MCS chain \texttt{next} pointer to refer to 
$N$, the thread waits for $P\rightarrow{}DWM$ to become non-0.  It then propagates that value 
into its own $N\rightarrow{}DMW$ field and finally sets $P\rightarrow{}Next = N$ in the normal MCS fashion.  
The MCS interlock between $N$ and $P$ -- where $P$'s next field is not yet set to $N$ -- is 
usually considered undesirable, as $P$, when calling unlock, may need to wait for 
$P\rightarrow{}Next$ to resolve to $N$.  We, however, leverage this property to our benefit,
``freezing'' $P$ briefly by inhibiting succession so we can safely read its $DWM$ field.  

Recapping, our approach propagates and \emph{pushes} the waitset forward, at unlock-time, 
from owner to successor.  And arriving threads in the lock operator \emph{pull} the 
DMW (hashCode) from their predecessor (or the mark word) into their MCS node. 
This allows the hashCode value to be read from the MCS tail element, which is referred to
directly by the mark word. 
All the key fields that otherwise reside in a classic ``fat'' object monitor are
now conveyed through the MCS chain.  We eliminate the centralized object monitor,
replacing it with the MCS chain, augmented to convey additional information.  

As a brief aside, it's worth pointing out that CJM is also amenable to designs that might
use just a single object header word.  See \url{https://wiki.openjdk.java.net/display/lilliput}.  
In this case the single 64-bit header word contains type information (an encoded reference to
class metadata), the hashCode, garbage collection``age'' bits, and low-order discriminated union ``tag'' bits that 
indicates if the header word is displace via synchronization activities.  
We expect that class information is effectively immutable.  

%% At this point we've fleshed out all the key operations, so it makes sense to examine
%% the advantages and disadvantage of the approach. 

\subsection{Example : CJM in use} 

\begin{figure}[h!]
\includegraphics[width=13cm,frame]{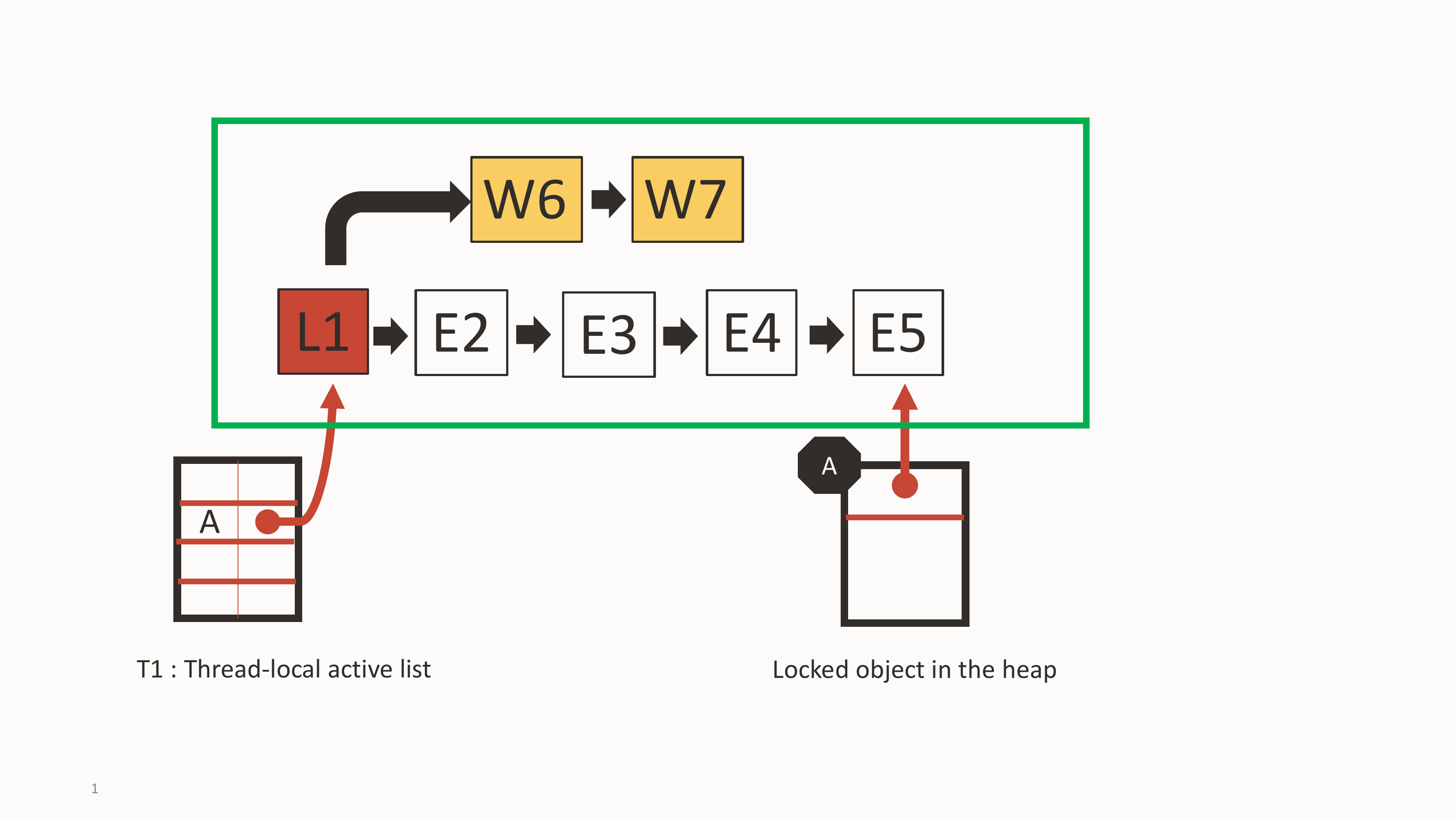}
\vspace{-8pt}      %% reduce whitespace between graph and caption
\caption{Example of CJM-MCS Queue}
\label{Figure:CJM}
\end{figure}

Figure-\ref{Figure:CJM} depicts a set of threads synchronizing on object $A$.  
$L1, E2, E3, E4, E5, W6, W7$ are MCS waiting elements (queue nodes) augmented for CJM. 
Each element is contributed by a distinct thread, $T1-T7$ respectively,  and resides 
on that thread's active list.  Eventually, those elements will cycle back to their thread's 
free list.  Thread $T1$ contributed $L1$, which is in \emph{LOCKED} state.
$L1$ resides at the head of the chain, thus $T1$ holds $A$.
$E1, E2, E3, E4, E5$ reside on the MCS chain and are in \emph{ENTRY} state,
and their associated threads are stalled waiting to acquire $A$.  
$A$'s mark word, which encodes the MCS tail, points to $E5$.
If a new thread arrives to acquire $A$, it enqueues after $E5$ and swings the MCS tail pointer. 
Elements $W6$ and $W7$ represent threads stalled in \texttt{wait} -- in \emph{WAITING} state. 
When notified, they will be removed from the waitset and appended to the tail of the MCS chain.  
\texttt{Notify} operates in constant time.  
When $T1$ calls \texttt{unlock($A$)}, the implementation locates $L1$ via its thread-local active list,
passes the waitset pointer (to $W6$ and $W7$) to its successor, $E2$, and finally passes
ownership to $E2$.  $E2$'s associated thread, $T2$, has a thread-local active list
that associates $A$ with $E2$ (not shown).  $T1$ then removes $L1$ from its 
active list and places $L1$ on its local free list, allowing the element to be reused
in future operations.  These last steps execute outside and after the critical section. 

\section{Variations} 

\BoldSection{Hashed External Waitset} Instead of propagating the waitset through 
the MCS chain, we can, instead, using 
the object's hashcode, hash into a shared array of waitset bucket elements, where
each bucket contains a lock and a pointer to set of MCS nodes associated with
active \texttt{wait} operations.  Hash collisions are possible, so the chain may
be mixed, with elements from different objects.  The bucket lock protects operations that
append, remove and pop elements from the chain.  This approach, which
significantly simplifies the implementation of \texttt{wait}, 
still allows wait morphing but at the expense of incurring extra locking operations.  
In particular, this approach obviates the need for the beta element and the special
waiting placeholder state.  

\BoldSection{Fissile Locks Optimization} As described above, CJM provide FIFO/FCFS succession
with direct handover, by virtue of its MCS lineage.  While fair, FIFO can suffer reduced throughput
compared to locks that allow so-called \emph{barging}, where departing threads release or renounce ownership
and arriving threads can \emph{pounce} on the lock and bypass threads that were already waiting.  
(The current HotSpot \texttt{synchronized} implementation, as well as, for instance, the default 
\texttt{pthread\_mutex} on linux, and the default \texttt{Java.util.Concurrent} \texttt{ReentrantLock}, 
allow unbounded bypass with unlimited starvation).  We could strike a compromise by using
\emph{Fissile Locks}\cite{dice2020fissile} which allow bounded bypass, trading improved throughput against
short-term unfairness.  To use Fissile locks we'd need to claim 2 additional bits in the mark word
to serve as the Fissile \emph{outer} lock, while the traditional MCS chain would act as 
the Fissile \emph{Inner} lock.  Fissile locks configured to use MCS as the inner lock require only
one MCS queue element per thread, regardless of the number of locks held.  And in fact that element
can be allocated on-stack as it is only needed while waiting.  But for use in CJM that particular
benefit of fissile locks is not available as we need the queue element to track held locks.  

\BoldSection{Mark-free Synchronization} Ignoring the hashCode value, it's entirely possible
to implement \texttt{synchronized} in a fashion that completely avoids using the mark word.  
Instead, we hash the object's address (or better, use its hashCode) into a table of buckets
where each bucket has a lock and a chain of \emph{prime} CJM MCS elements.  A prime element is the 
nominal head of the MCS chain for a given object.  Prime elements, each associated with
a distinct object, form the spine of the list, and individual entrysets and waitsets are
linked from each prime, as ribs. 

\Invisible{spine-and-rib list formation} 
\Invisible{Slow-down : collisions and bucket-lock acts as an inner lock}  
\Invisible{Mark-free AKA no-touch} 

We've implemented the equivalent of this approach in LD\_PRELOAD libraries
that interpose on the \texttt{pthread} synchronization primitives, and, not
surprisingly, found that latency for uncontended lock and unlock increases 
and scalability is impacted.  

If we were willing to take 2 bits from the other header word, and use the IBM 2-bit 
hashCode technique, putting aside the issue of GC age bits, we could eliminate the mark word,
but the price to pay in reduced synchronization performance is non-trivial. 

\Invisible{Coupling; dependencies; interlace} 

\section{Advantages} 

\begin{enumerate} 

\item Compared to the existing implementation, simpler and less code. 

\item Very little coupling between safepoints, garbage collection, and the synchronization subsystem.

\item No inter-thread migration of monitors or other elements

\item Tightly bounded footprint, with no need to scavenge monitors at safepoints.

\item MCS queue elements don't require type-stable memory providing simple
lifetime and lifecycle for queue nodes.

\item FIFO fair lock admission with more predictable performance.

\item All or most of the algorithm could be transliterated to Java instead of C++,
allowing a mostly-in-java implementation except for calls to native \texttt{park} and \texttt{unpark},
although some new unsafe helpers and GC accomodation would be needed to deal with the mixed-type mark word 
\footnote{This has been a long sought-after goal with attempts made in various managed runtimes.  
See also \url{https://wiki.se.oracle.com/display/JPG/Java+Object+Monitors+in+Java}}. 
The result \emph{may} be more friendly to project loom than the current implementation. 
Queue elements would then become first-class Java objects.  

\Invisible{
An excellent anti-pattern is IBM's ``Jikes'' research JVM. 
A good example of getting it right is Microsoft's C\# implementation of 
language-level synchronization primitives via unsafe blocks, which provide a sufficiently 
rich dialect in which to implement ``system'' constructs.  Where, in HotSpot, our implementation
might span Java library code, C2 ``ideal node'' graph dialect, and native C++, the C\# runtime 
can package the equivalent implementation in one source file, skirting Conway's Law}.  

\item Loom may benefit from succession by direct handoff.  

\item Eliminates the objectMonitor construct, presenting an opportunity to remove code.

\item Eliminates stack locking mode, reducing all synchronization operations
to just one flavor.  

\item Simplified mark word encoding with only 1 tag bit required

\item Still provides wait morphing.

\Invisible{burried under; subsumed into;} 

\item All state transitions are subsumed into \texttt{park}.  

\item Uncontended performance (latency) is on par what we have with 
what we have in stack locking in the existing implementation. 

\item Lock-unlock performance is about the same as classic MCS, as we're not accessing
any additional cache lines (no additional concurrency traffic) despite the slightly more
involved paths. 

\item As a stretch goal we easily adapt the lock to be NUMA-aware via CNA

\item As we've seen with CNA, etc, we can be quite flexible regarding the admission order,
and not necessarily constrained to FIFO.  

\item  MCS is well understood and widely used, and CJM is really just a variation 
 on MCS. 

\item Eliminates spurious wakeups in \texttt{wait}.  

\item No tunable performance ``knobs'' or opaque heuristics

\end{enumerate}

\section{Disadvantages} 

\begin{enumerate} 

\item Expunging objectMonitors and stack locks, while an admirable long-term goal to reduce
technical debt, is a heavy lift, as those constructs have tendrils everywhere. 
Early prototypes could leave the defunct structures in place, however. 
Specifically, as an interim measure, it should be possible to leave a degenerate transitional \texttt{objectMonitor} 
class in place which contains just a references to the associated object.  

%% \item The algorithm is fairly ornate but that may be a fact of life 
%% given the constraints and the overloaded mark word

\item FIFO (and more generally, succession by direct handoff) provide predictable
performance \footnote{Helpful for SLAs.} but throughput and scalability is generally 
less than that achieved by schemes that allow barging (either unbounded or bounded bypass) 
\cite{dice2020fissile}.  This reflects the classic fairness vs throughput trade-off. 

\item Under preemption, FIFO performance can be problematic as we risk handover to 
preempted threads, stalling progress.  A viable work-around is to divert \emph{excess} threads
not needed to maintain saturation of the lock to side lists 
\cite{arxiv-Malthusian,eurosys17-dice,arxiv-GCR,EuroPar19-GCR}.  

\item FIFO and spin-then-park waiting -- with simplistic fixed bounded spin durations -- 
don't usually work well together, as the immediate
successor is likely to have waited the longest and, for constant spin durations, thus 
be most likely to have already parked, impacting lock hand-over latency.   
This may manifest as a surprising and abrupt non-linearity in performance 
when the system rapidly undergoes a phase change where most successor are 
found to be parked.  Techniques such as culling parked threads from the chain, 
and anticipatory wakeup can be palliative. 

\item Auxiliary services such as JVMTI/DI, etc would be impacted and likely require attention 

\item It is easy to tell if the current thread $T$ holds $O$, but harder, given just $O$, to 
determine the owner of $O$ or if some arbitrary other thread $S$ holds $O$.  
This may impact some support code.  Relatedly, on-the-fly deadlock detection is more difficult.

\item Once an implementation has switched to a FIFO it can be difficult to go back as 
applications may inadvertently come to depend on the admission policy.  
We've encountered producer-consumer idioms, for instance, that poll or 
busy-wait via a lock, and one flavor can happen to be starved out.  
So switching to FIFO might be a one-way transition and limit our latitude 
in future designs. 

\item While the algorithm proper is relatively self-contained, the changes
would likely be invasive and extensive. 

\item Arrivals mutate the object's mark word, as it serves as a pointer
to the MCS tail element.  This may result in additional false sharing
between accesses to the mark and accesses to the rest of the object.  

\end{enumerate} 

%% \section{Conclusion} 

\appendix

\section{Monitor deflation and inflation strategies and lifecycle} 
\label{Appendix:lifecycle} 

\Invisible{Characteristics and Properties : 
inflation; deflation; accumulation; flow and migration; protocol and dance;
} 

%% LISTS: https://en.wikibooks.org/wiki/LaTeX/List_Structures
\ListProperties(Hide=100, Space1*=.1cm, Space2*=.1cm, Hang=true, Progressive=3ex, Style*=-- ,
Style2*=$\bullet$ ,Style3*=$\circ$ ,Style4*=\tiny$\blacksquare$ )

\Boldly{Legacy HotSpot} 
\begin{easylist} 
& \textbf{Inflation: } On lock contention, \texttt{wait}, etc.  Monitors are
allocated one-at-a-time from a central pool, under a lock.  
& \textbf{Deflation: } Lazily at safepoint time.  This policy increases
safepoint durations because of the scavenging phase, which is effectively serial, 
and degrades overall performance.  There is no natural back-pressure to trigger scavenging.  
& \textbf{Migration: } None
& \textbf{Accumulation :} Rampant monitor accumulation is possible and has been observed in
the wild, with associated bug reports filed.  The issue
of memory pressure is exacerbated by the fact that monitors reside in type-stable memory.  
& \textbf{Access protocol: } Simple with no overheads as the object:monitor relationship is stable
except over safepoints.  
& \textbf{Lifecycle :} Threads allocate \texttt{objectMonitors} from a central pool
one-at-a-time.  Eventually the safepoint scavenger determines that a monitor is idle,
deflates the object, and places the monitor back on the central free list.  
\end{easylist} 

\Invisible{
\noindent\rule{\textwidth}{1pt}
\RedZone{Caveat: I'm not as familiar with the changes either already in or planned for ``Modern HotSpot''
so this section may include inaccuracies or critical omissions.  Clarification, corrections and comments are welcome.\\} 
\noindent\rule{\textwidth}{1pt}
} 

\Boldly{Modern HotSpot}
\begin{easylist} 
& \textbf{Inflation: } On lock contention, \texttt{wait}, etc.  Monitors are
allocated one-at-a-time from a central pool, under a lock.  
& \textbf{Deflation: } Deferred.  Deflation and scavenging runs outside safepoints and is
performed by a dedicated \texttt{MonitorDeflationThread}, improving safepoint pause
times. 
& \textbf{Migration: } None
& \textbf{Accumulation :} Improved relative to Legacy HotSpot, above, as deflation and 
safepoints are decoupled and we can run deflation more frequently to keep the number
of extant circulating monitors in check.  A lingering concern is rate-mismatch between
inflation, and production of idle monitors, and the system's ability to
to deflate and recover those monitors.  That is, can deflation keep up with inflation. 
& \textbf{Access protocol: } Complex with overheads impacting normal lock and
unlock operations.  Races are possible between threads trying to reach a monitor
from the mark word, and concurrent deflation.  Uses a \texttt{SENTINEL} protocol
to detect and recover from the race.  
\end{easylist} 

\Boldly{ExactVM} 
\begin{easylist} 
& \textbf{Inflation: } Aggressive and early -- on initial lock operation
& \textbf{Deflation: } Early -- ASAP at runtime by normal mutator threads
when the unlock operator detects the monitor is idle with no threads in the waitset or entryset.
& \textbf{Migration: } Vulnerable -- mitigations necessary 
& \textbf{Accumulation :} Constrained 
& \textbf{Access protocol: } Complex with overheads impacting normal lock and
unlock operations.  Races are possible between threads trying to reach a monitor
from the mark word, and concurrent deflation.  Uses the \emph{metalock} to prevent
the race. 
\end{easylist} 

\Boldly{CJM}
\begin{easylist} 
& \textbf{Inflation: } Immediate -- all lock operations contribute a queue element to
the chain, which is recovered in the corresponding unlock operation.  
& \textbf{Deflation: } ASAP
& \textbf{Migration: } None.  
CJM avoids imbalanced flow that can occur in other implementatons, where one thread 
inflates and other subsequently deflates.  In addition, CJM avoids the use of reference counts 
as there is no central per-lock structure requiring inflation or deflation.   
& \textbf{Accumulation :} Minimal 
& \textbf{Access protocol: } Simple
& \textbf{Remarks: } No central monitor pool or locking thereof
\end{easylist}

\section{A quick overview of MCS locks} 
\label{Appendix:MCS} 

\BoldSection{MCS: } 
The \emph{MCS lock} \cite{tocs91-MellorCrummey}, is the usual alternative to simple
test-and-set-based locks, or ticket locks, performing better under high contention, but
also having a more complex path and often lagging behind simple locks under no or light contention.
In MCS, arriving threads use an atomic \texttt{SWAP} operation to append an element
to the tail of a linked list of waiting threads,
and then busy wait on a field within that element, avoiding global spinning as found in test-and-set locks.
The list forms a queue of waiting threads.  
The lock's tail variable is explicit and the head -- the current owner --
is implicit. When the owner releases the lock it reclaims the element it
originally enqueued and sets the flag in the next element, passing ownership.
%%  using \emph{local spinning} where at most one thread waits on a given location at any time. 
To convey ownership, the MCS unlock operator must identify the successor, if any, and then
store to the location where the successor busy waits.
The list forms a multiple-producer-single-consumer (MPSC) queue where any thread can enqueue but
only the current owner can dequeue itself and pass ownership.  
The handover path is longer than that of test-and-set locks and accesses more distinct
shared locations.  

MCS uses so-called \emph{local waiting}\cite{topc15-dice}  where at most one thread is waiting on a given
location at any one time.  As such, an unlock operation will normally
need to invalidate just one cache line -- the line underlying the flag where the successor busy waits --
in one remote cache.  (Lock algorithms that use local spinning are also
trivially easy to convert to blocking via park-unpark).  
Under contention, the unlock operator must fetch the address of the successor
element from its own element, and then store into the flag in the successor's element,
accessing two distinct cache lines, and incurring a dependent memory access to reach the successor.
Absent contention, the unlock operator uses an atomic compare-and-swap (CAS) 
operator to try to detach the owner's element and set the tail variable to \texttt{null}. 

MCS locks provide strict FIFO order.  They are also compact, with the lock
body requiring just a pointer to the tail of the chain of queue elements.  

One MCS queue element instance is required for each lock a thread currently holds, and
an additional queue element is required while a thread is waiting on a lock.
Queue elements can not be shared concurrently and can appear on at most one queue
-- be associated with at most one lock -- at a given time.
The standard POSIX \texttt{pthread\_mutex\_lock} and \texttt{pthread\_\allowbreak{}mutex\_\allowbreak{}unlock} 
operators do not require scoped or lexically balanced locking.  
As such, queue element can not be allocated on stack.   
Instead, MCS implementations that expose a standard POSIX interface will typically allocate elements 
from thread-local free lists, populated on 
demand \footnote{We note that the MCS ``K42'' variant \cite{K42,Scott2013} allows queue elements to 
be allocated on stack -- they are required only while a thread waits -- 
but at the cost of a longer path with more accesses to shared locations.}. 

\Invisible{MCS requires the address of queue element inserted by the owner to 
be passed to the corresponding unlock operator, where it will be used to identify 
a successor, if any.} 

The standard POSIX interface does not provide any means to pass information 
from a lock operation to the corresponding unlock operator.  As such, the address of 
the MCS queue element inserted by the owner thread is usually recorded in an additional field in the lock 
instance so it can be conveyed to the subsequent unlock operation to identify the successor, if any.
That field is protected by the lock itself and accessed within the critical section.  
Accesses to the field that records the owner's queue element address may themselves generate 
additional coherence traffic, although some implementations may avoid such accesses to 
shared fields by storing the queue element address in a thread-local associative structure that maps 
lock addresses to the owner's queue element address.

\bibliography{cjm}

\end{document}